\documentclass[preprint,showpacs,preprintnumbers,amsmath,amssymb]{revtex4}
\usepackage{graphicx}
\newcommand{\be}{\begin{equation}}

\newcommand{\ee}{\end{equation}}
\newcommand{\bea}{\begin{eqnarray}}
\newcommand{\eea}{\end{eqnarray}}
\newcommand{\beas}{\begin{eqnarray*}}
\newcommand{\eeas}{\end{eqnarray*}}

\begin{document}

\title{Coulomb proximity decay: A clock for measuring the particle emission time scale}

\author{S. Hudan}
\author{A.S. Botvina}
\altaffiliation[]{On leave from Institute for Nuclear Research, Moscow, Russia}
\author{R. Alfaro}
\author{L. Beaulieu}
\author{B. Davin}
\author{Y. Larochelle}
\author{T. Lefort}
\author{V.E. Viola}
\author{H. Xu}
\author{R. Yanez} 
\author{R.T. de Souza} 
\affiliation{
Department of Chemistry and Indiana University Cyclotron Facility \\ 
Indiana University, Bloomington, IN 47405} 

\author{T.X. Liu}
\author{X.D. Liu}
\author{W.G. Lynch}
\author{R. Shomin}
\author{W.P. Tan}
\author{M.B. Tsang} 
\author{A. Vander Molen} 
\author{A.Wagner}
\author{H.F. Xi}
\affiliation{
National Superconducting Cyclotron Laboratory and Department of
Physics and Astronomy \\ 
Michigan State University, East Lansing, MI 48824}

\author{R.J. Charity}
\author{L.G. Sobotka}
\affiliation{
Department of Chemistry, Washington University, St. Louis, MO 63130}

\date{\today}

\begin{abstract}
A new method to examine the time scale of particle emission 
from hot nuclei is explored. 
Excited projectile-like and 
target-like fragments decay as they separate 
following a peripheral heavy-ion collision. Their mutual Coulomb influence 
results in an anisotropic angular distribution of emitted particles, 
providing a measure of the particle emission time scale. 
Predictions of a schematic evaporation model are presented and compared to 
experimental data.

\end{abstract}

\maketitle

While the decay of an isolated nucleus is well described by 
a statistical emission process\cite{Ericson60}, the decay in the presence
of an external inhomogeneous field has only recently 
been considered\cite{Botvina99}. 
Neutron stars with strong gravitational fields and heavy-ion collisions
with strong Coulomb fields are environments in which such non-isolated
decay may occur.
In this Letter, we consider the Coulomb influence on the decay of a hot 
nucleus formed in a 
heavy-ion collision.
For a nucleus at modest excitation (E$^*$/A$\ge$2 MeV), statistical
model calculations predict decay on a relatively
fast time scale, $t$$\le$100 fm/c\cite{Charity00}. 
Directly measuring such short lifetimes 
is difficult\cite{Hilscher} and 
is presently limited to a few approaches: 
two-fragment correlations\cite{Trockel87,Cornell95,Beaulieu00} and
near scission particle emission\cite{Hinde92}.
While two-particle correlations measure the time between successive 
emissions, and work at high excitation, near scission emission is 
restricted to relatively low excitation and the long time scale 
associated with fission. Other approaches such as the crystal blocking 
technique are sensitive to even longer time scales
\cite{Goldenbaum99}.
We describe a new approach to measuring the time scale of 
particle emission from hot nuclei, namely the angular anisotropy associated 
with Coulomb proximity decay.

For light fragments, Coulomb proximity decay has previously been 
used to explain the correlation
between relative velocities of the decay products and their relative
orientation\cite{Charity01} in terms of tidal forces\cite{Charity92}. 
For heavy projectile-like fragments, from in-plane angular distributions, 
the time between successive binary splittings\cite{Casini93,Durand95} 
has been deduced.
In this analysis, we report
for the first time on the influence of the external field on the 
probability for binary decay {\it and} the resulting angular distribution
of emitted particles.
Peripheral collision of two intermediate energy 
(E/A = 20-100 MeV) heavy-ions results in 
excited projectile-like and 
target-like fragments (PLF$^*$ and TLF$^*$ respectively) which de-excite 
as they separate with particle emission modified by their mutual 
Coulomb interaction. 
For a fixed separation distance between the PLF$^*$ 
and TLF$^*$, and an asymmetric split of the PLF$^*$, 
emission of the smaller nucleus towards the TLF$^*$ is favored 
due to the reduced Coulomb 
interaction between the three bodies\cite{Botvina99,Botvina01}. 
During the split, while the center-of-mass of the TLF$^*$-PLF$^*$ system does 
not change, the center-of-mass of the PLF$^*$ decay products can fluctuate. 
Although the
influence of the external Coulomb field on the instantaneous breakup of 
a highly excited PLF$^*$ has been reported\cite{Botvina99,Botvina01}, 
we now examine 
the proximity influence on successive binary emissions and the resulting 
'clock' of statistical emission.

To examine the main features of Coulomb proximity decay
we have constructed a model in which 
the PLF$^*$ with (Z,A), characterized by a spin (J) and excitation (E$^*$/A),
moves away from the TLF$^*$ with a velocity, V.
For simplicity, the TLF$^*$ is represented by a point source which does not
undergo decay.
Starting at an initial separation distance, the de-excitation 
of the PLF$^*$ {\it via} sequential binary decays 
of light particles ($n$, Z$\le$2) and 
heavy clusters (up to $^{18}$O)
is calculated using a 
Weisskopf approach\cite{Weisskopf37}. 
In this model\cite{Botvina87}, the decay width for the emission of a particle $j$ in 
the excited state $i$ 
from a nucleus $(Z,A)$ is given by:
\be \label{eq:eva}
\Gamma_{j}^{i}=\int_{0}^{E_{AZ}^{*}-B_{j}-\epsilon_{j}^{(i)}}
\frac{\mu_{j}g_{j}^{(i)}}{\pi^{2}\hbar^{2}}\sigma_{j}(E)
\frac{\rho_{f}(E_{CN}^{*}-B_{j}-E)}{\rho_{CN}(E_{CN}^{*})}EdE
\ee
Here $\epsilon_{j}^{(i)}~(i=0,1,\cdots,n)$ 
is taken for the ground and all particle-stable excited states
of the fragment $j$, and
$g_{j}^{(i)}=(2s_{j}^{(i)}+1)$  is  the
spin degeneracy factor of the $i$-th excited  state,
$\mu_{j}$ and $B_{j}$ are corresponding reduced mass and separation energy,
$E_{CN}^{*}$ is the excitation energy of the initial compound nucleus 
(PLF$^*$), and
$E$ is the kinetic energy of an emitted particle in the center-of-mass 
of the emitting system. The level densities of the initial compound $(A,Z)$ and final $(A_{f},Z_{f})$
residual nuclei, $\rho_{CN}$ and $\rho_{f}$ are calculated using the 
Fermi-gas formula 
$\rho(E)\propto exp\left(2\sqrt{aE}\right)$ with the level density 
parameter $a \approx 0.125A$MeV$^{-1}$.
In the present 
work we have parametrized the cross-section 
as $\sigma_{j}(E)=\pi R_{fj}^{2}(1-U_{c}/E)$, 
where $U_{c}$ is the Coulomb barrier for fragment emission and 
$R_{fj}$=$R_{f}$+$R_{j}$, $R_{f}$=$r_{n}A_{f}^{1/3}$, 
$R_{j}$=$r_{n}A_{j}^{1/3}$, with $r_{n}=1.5$fm.

We modified this model by taking into account the influence of 
the external Coulomb field on the decay width.  
The daughter residue and emitted fragment are assumed to be touching spheres 
and are placed  
within a sphere of radius $R_{j}+R_{f}$ with center located at 
the center-of-mass of the decaying
parent (PLF$^*$) nucleus. The decay configuration within the sphere is impacted  
by the change in the Coulomb energy of the three-body system: 
\be
U_{c}=\frac{Z_f*Z_j}{R_{fj}}+ 
\frac{Z_{TLF*}Z_{j}}{R_{TLF*j}}+\frac{Z_{TLF*}Z_{f}}{R_{TLF*f}}-\frac{Z_{TLF*}Z}{R_{TLF*CN}}
\ee
 where 
$Z_{TLF*}$ is the charge of the TLF$^*$, and $R_{TLF*j}$, $R_{TLF*f}$, 
$R_{TLF*CN}$ 
are distances from the TLF$^*$ to the emitted particle, 
residual nucleus and compound nucleus, respectively. 
The probability of fragment emission 
can be found by averaging eq. (\ref{eq:eva}) over all coordinates of 
fragments. Since the Coulomb barrier is lower when the smaller fragment is 
emitted in the direction of the TLF$^*$, the 
resulting fragment emission is anisotropic in the PLF$^*$ frame. The angular
momentum of the PLF$^*$ was included using a standard approach
\cite{Botvina01,Charity88}.

The full width for evaporation is determined by summing up all emission 
channels: $\Gamma=\sum \Gamma_j^i$. The mean time for an emission 
step is given by $\tau= \hbar / \Gamma$. 
The PLF$^*$ 
is assumed to have a lifetime, $t$,  
with a distribution $exp(-t/\tau)$. 
Until the decay occurs, the PLF$^*$, TLF$^*$, and 
all charged particles 
propagate along Coulomb trajectories. Successive 
binary emissions, conserving energy and momentum, are calculated 
until the excitation energy is below the particle emission threshold.

For the results presented we assumed 
the
PLF$^*$ had Z=38, A=90 and was located at an initial distance of 30 fm from the
TLF$^*$ with 
spin J=10$\hbar$ or J=40$\hbar$. This distance is compatible with
an equilibration time of $\approx$100 fm/c following the collision in agreement
with dynamical simulations\cite{Bondorf95}. 
The TLF$^*$ was assumed to be a point source with 
Z=42, A=114, V=0.2728c relative to the PLF$^*$ and interacted
{\it via} the Coulomb interaction. 

\begin{figure}[t] 
\includegraphics [scale=0.40]{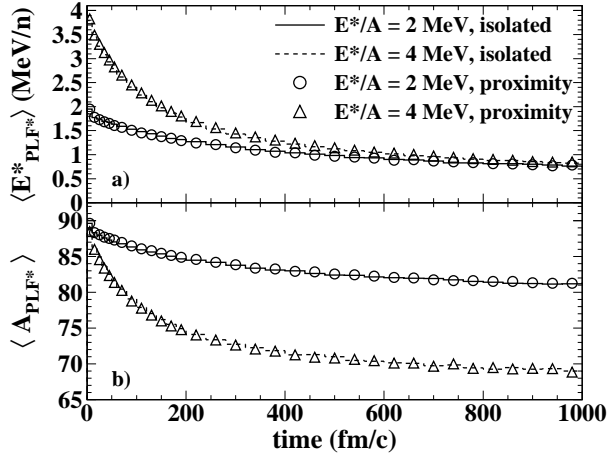}%
\caption{\label{fig:fig1}
Average excitation energy and mass number
as a function of time for  
a nucleus decaying in an external field and an isolated nucleus.
} 
\end{figure}

Displayed in Fig.~\ref{fig:fig1} is the influence of the Coulomb proximity effect 
on the de-excitation of the PLF$^*$ with 
J= 10$\hbar$ . 
The decay of an isolated PLF$^*$, not subject to the 
Coulomb proximity effect, is presented as a reference.
The essentially identical behavior in both $\langle$E$^*$$_{PLF*}$$\rangle$ and
$\langle$A$_{PLF*}$$\rangle$ as a function of time
for both the isolated and proximity cases, 
indicates that for these initial conditions, the average rate of 
particle emission is comparable. 
For smaller initial PLF$^*$-TLF$^*$ distances or larger TLF$^*$ 
atomic number, however, the proximity influence can alter the particle 
emission rate as 
compared to the isolated case.
The rapid de-excitation of the PLF$^*$ is clearly evident for $t$$\le$250 fm/c,
corresponding to a PLF$^*$-TLF$^*$ distance of $\approx$70fm. 
Associated with the rapid 
decrease in excitation of the PLF$^*$ is the corresponding decrease in the mass
of the PLF$^*$, (Fig.~\ref{fig:fig1}b). 
For an initial excitation of E$^*$/A= 4 MeV, by $t$=250 fm/c 
approximately 25\% of the mass of the PLF$^*$ has been emitted.

\begin{figure}[t]
\includegraphics [scale=0.40]{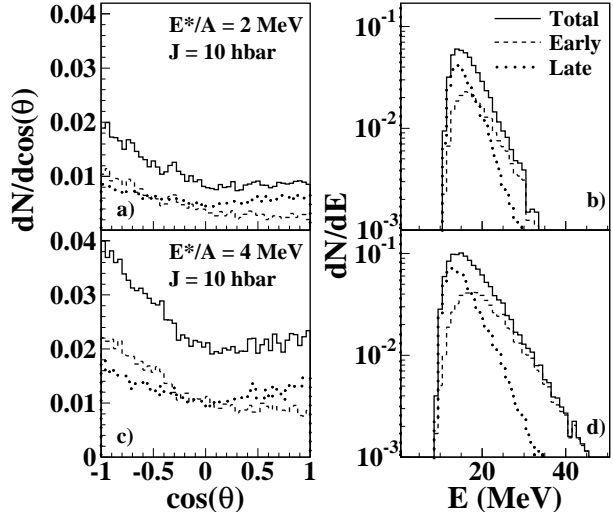}%
\caption{\label{fig:fig2}
Angular distributions and kinetic energy spectra of $\alpha$ 
particles in the PLF$^*$ frame selected on emission time.} 
\end{figure}

Although the average excitation of the PLF$^*$  
at a given time, under these assumptions,
is not appreciably affected by the Coulomb proximity, the angular
distribution of emitted particles is 
impacted by the Coulomb field of the target,
as shown in Fig.~\ref{fig:fig2} for $\alpha$ particles 
(E$^*$/A = 2 and 4 MeV and J = 10$\hbar$).
For all cases, the angular distribution is peaked near cos($\theta$)=-1, 
corresponding to emission in the direction of the TLF$^*$. For the 
conditions calculated, emission towards 
the TLF$^*$, i.e. cos($\theta$)$\approx$-1.0,  is enhanced in yield by 
a factor of 2 relative to cos($\theta$)$\approx$0.
To understand better the
behavior of proximity emission we have selected on decays 
that occur at a 
distance $\le$70 fm (i.e. $\approx$250 fm/c) and refer to these decays 
as early decays. Decays which occur at larger distances 
experience a negligible
influence of the TLF$^*$ Coulomb field and are referred to as 
late decays. 
While the late 
decays (dotted line) exhibit a symmetric and 
relatively isotropic distribution, the 
pronounced peak in the total angular distribution is associated with
the early decays (dashed line) which are focused towards 
the TLF$^*$ by the proximity Coulomb effect. The magnitude of the asymmetry
for these early decays,
can be assessed, to first order, by comparing the yield at 
cos($\theta$)=-1 to cos($\theta$)=+1. At E$^*$/A=2 MeV an asymmetry 
of $\approx$4 is observed.

The impact of the Coulomb proximity effect on the kinetic energy spectra 
of emitted particles is also shown in Fig.~\ref{fig:fig2}.
In this figure, it is evident that, as expected\cite{huizenga89}, the tail
of the $\alpha$ particle kinetic energy distribution is preferentially 
populated by early emissions (dashed line). 
From the angular distributions presented in Fig.~\ref{fig:fig2} 
it is clear that these emissions 
occur in the proximity of the TLF$^*$. Comparison of Fig.~\ref{fig:fig2}b 
and \ref{fig:fig2}d shows that with increasing excitation, 
late emission (dotted line) provides an increasing contribution 
to the region of the Coulomb peak. For $\alpha$ with E$\le$25 MeV, the magnitude
of the increase for the late emission is $\approx$6\%.  
This result is understandable 
as with increasing excitation, an increasing amount of
excitation energy remains at a PLF$^*$-TLF$^*$ distance of 70 fm 
where the proximity effect is significantly reduced. The energy-angle
correlations introduced by the Coulomb proximity effect present 
interesting opportunities in studying the de-excitation cascade of a 
hot nucleus. The angular asymmetry of $\alpha$ particles resulting from 
proximity emission is related to the time spent by the PLF$^*$ in the 
vicinity of the TLF$^*$ and thus forms a 'clock' of
the statistical emission time.

\begin{figure}[t] 
\includegraphics [scale=0.30]{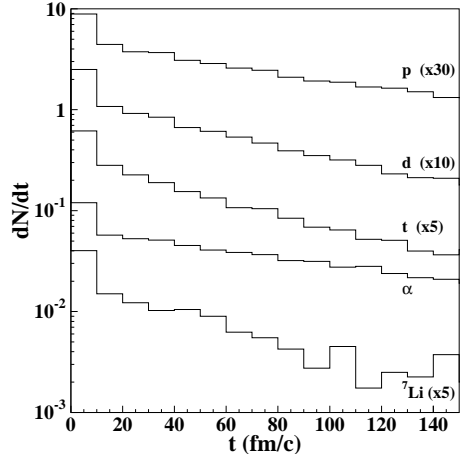}%
\caption{\label{fig:fig3}
Emission time distributions of emitted particles from the PLF$^*$.} 
\end{figure}

Another interesting aspect of the Coulomb proximity effect is that 
different emitted particles are sensitive to different portions of the
de-excitation cascade. 
Depicted in Fig.~\ref{fig:fig3} is the dependence of 
emission probability on time for p,d,t,$\alpha$ and $^7$Li for 
E$^*$ = 4 MeV and J=10$\hbar$.
While the emission time distributions exhibited of  
$\alpha$ particles, for example is relatively flat decreasing by only a factor 
of 3 over the time interval displayed, 
other particles e.g. t, $^7$Li exhibit steeper distributions. In  the case of 
both t and $^7$Li, a decrease of more than one order of magnitude is observed. 
This behavior can be 
understood in terms of the Coulomb barrier and binding energy  
relative to 
the available excitation of the PLF$^*$.  Particles such as $^7$Li therefore
sample the earliest 
portion of the 
de-excitation cascade and thus exhibit 
an enhanced sensitivity to proximity emission. 
Comparison of the angular 
asymmetry of particles with different sensitivities can therefore 
provide multiple 
'clocks' which probe the statistical decay of a hot nucleus.
Naturally, the Coulomb field of a particle emitted early also influences 
subsequent emissions. Consequently, this work represents
the first step towards a theory of {\it interacting sequential decays}. As 
such it represents a significant departure from Bohr's independence 
hypothesis that is traditionally invoked.

\begin{figure}[t] 
\includegraphics [scale=0.40]{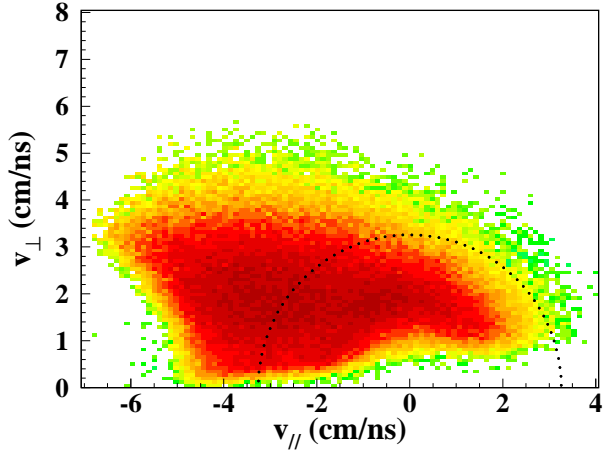}%
\caption{\label{fig:fig4}
Invariant cross-section map of $\alpha$ particles in the PLF$^*$ frame for the 
reaction $^{114}$Cd + $^{92}$Mo at E/A = 50 MeV.
} 
\end{figure}

In order to examine whether the asymmetries predicted by this schematic model 
are manifested in experimental data with comparable magnitude, 
we have compared the model predictions 
to data from the
reaction $^{114}$Cd + $^{92}$Mo at E/A=50 MeV. 
In this experiment, detection of
a PLF (15$\le$Z$\le$46) at
very forward angles (2.1$^{\circ}$$\le$$\theta_{lab}$$\le$4.2$^{\circ}$) 
together with
measurement of the associated charged particles emitted at 
larger angles 
(7$^{\circ}$$\le$$\theta_{lab}$$\le$58$^{\circ}$) 
allow us 
to focus on particles emitted from the PLF$^*$\cite{Yanez03}. 
The invariant cross-section map shown in Fig.~\ref{fig:fig4}, exhibits
an essentially circular ridge of yield centered on the PLF$^*$ velocity. 
This ridge has a radius 
corresponding to the 
Coulomb repulsion between the emitted $\alpha$ and PLF$^*$ residue 
suggesting that
$\alpha$ particles along this ridge are associated with emission from the 
PLF$^*$. Selecting $\alpha$ particles on the ridge does not ensure 
statistical emission as dynamical
processes may also contribute
at the most backward angles\cite{Bocage00}.
Also evident in Fig.~\ref{fig:fig4} is a peak at V$_{\parallel}$=-4 cm/ns and 
V$_{\perp}$=2 cm/ns, attributable to mid-velocity 
emission\cite{plagnol00,Gingras02} that 
represents a background for the statistical component considered.

\begin{figure}[t!] 
\includegraphics [scale=0.40]{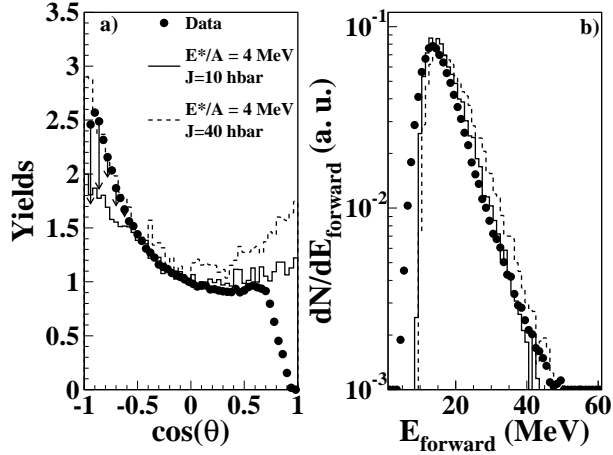}%
\caption{\label{fig:fig5}
Panel a: Comparison of angular distribution of $\alpha$ particles from the 
reaction $^{114}$Cd + $^{92}$Mo at E/A = 50 MeV with 
E$_{\alpha}$$\le$22 MeV in the PLF$^*$ frame with results of the
proximity decay model.
Panel b: Kinetic energy spectra in the PLF$^*$ frame of $\alpha$ particles
emitted forward of the PLF$^*$.
} 
\end{figure}

In Fig.~\ref{fig:fig5}a the angular distribution of $\alpha$ particles with 
E$_{\alpha}$$\le$22 MeV, i.e. along the ridge in Fig.~\ref{fig:fig4}, 
is displayed normalized to the yield at $\theta$=90$^\circ$. 
The data are backward peaked with an enhancement of a factor 
of $\approx$2. The observed $\alpha$ particle yield at
forward angles, cos($\theta$)$\ge$0.65, and very backward angles,
cos($\theta$)$\le$-0.9, is supressed due to the 
finite acceptance of the experimental setup. 
We have corrected for the mid-velocity emission contribution to the 
yield with 
E$_{\alpha}$$\le$22 MeV 
by examining the kinetic energy distributions in the 
PLF$^*$ frame selected on emission angle, 
as indicated by the representative arrows 
in Fig.~\ref{fig:fig5}.
For cos($\theta$)$\ge$-0.7 
this contribution is negligible while for more backward angles  
a correction of upto 25$\%$ is observed.
The excitation deduced by calorimetry, using forward emission and assuming
isotropy, for the data shown in Fig~\ref{fig:fig4}
and \ref{fig:fig5} is $\approx$2.3 MeV/A\cite{Yanez03}.  
The measured angular 
distribution is reasonably described by the 
model presented with E$^*$/A = 4 MeV. 
The discrepancy between the experimentally deduced excitation and the initial
excitation used in the model can be understood by 
the rapid cooling evident in Fig.~\ref{fig:fig1}. 
Inclusion of angular momentum at this excitation results in a
more peaked angular distribution for cos($\theta$)$\le$-0.7. 

The kinetic energy spectrum of $\alpha$ particles emitted in the forward 
direction is shown in Fig~\ref{fig:fig5}b. 
The experimental distribution is remarkably well 
described by the model, despite its simplicity. Increased angular 
momentum acts to broaden the peak of the distribution, as well as slightly 
extend the high energy tail of the distribution. While the exponential slope 
of the kinetic energy spectrum is somewhat better described by the 10$\hbar$
case, the width of the Coulomb peak is better described by the 40$\hbar$ 
case, perhaps suggesting that the experimental data is intermediate 
between these two cases, in agreement with the angular distributions shown in
Fig.~\ref{fig:fig5}a.

In summary, we have explored how Coulomb proximity decay can impact
the statistical binary decay of a hot nucleus. 
As PLF$^*$ and TLF$^*$ separate following a peripheral heavy-ion collision,.
while the emission rate of particles from the PLF$^*$ may be  
essentially unaffected, the angular distribution of emitted particles shows
a pronounced effect.
Emission of particles as the PLF$^*$ and TLF$^*$ separate
occurs preferentially in the direction of the TLF$^*$ resulting in an 
asymmetric angular distribution that reflects the amount of emission  
occuring in the vicinity of the TLF$^*$. 
This angular asymmetry for different emitted particles together with their 
associated kinetic energy spectra 
provides a clock of the emission timescale. 
Experimental data exhibit an asymmetry of comparable magnitude 
suggesting the applicability of this technique. 
Both the sensitivity 
of this clock and the deconvolution of dynamically emitted mid-rapidity
particles can be assessed by modifying the magnitude of the external field
by changing the target, modifying the velocity with which the PLF$^*$
and TLF$^*$ separate by changing the incident energy, or altering the 
excitation of the PLF$^*$ by selecting different velocity 
dissipation\cite{Yanez03}.

\begin{acknowledgments}
	We would like to acknowledge the 
valuable assistance of the staff at MSU-NSCL for
providing the high quality beams which made this experiment possible. 
This work was supported by the
U.S. Department of Energy under DE-FG02-92ER40714 (IU), 
DE-FG02-87ER-40316 (WU) and the
National Science Foundation under Grant No. PHY-95-28844 (MSU).\par
\end{acknowledgments}

\bibliography{proximity.bib} 

\end{document}